# Frozen time in hyperbolic spacetime motion

**Arne Bergstrom**

Paper      pages    1 – 5

FAQ        pages    6 – 17

# Frozen time in hyperbolic spacetime motion

Arne Bergstrom

B&E Scientific Ltd, Seaford BN25 4PA, UK

E-mail: arne.bergstrom@physics.org

**Abstract**
According to the Lorentz transformation and clearly seen from the Minkowski diagram, hyperbolic spacetime motion of a test object relative to a stationary reference frame can be performed in a specific way such that time becomes frozen in the moving frame of the test object. In that case, time can be arranged to become frozen even at moderate relativistic velocities, in contrast to the minute traditional relativistic time dilation at such velocities. An appendix gives a simple illustration in Minkowski form of how time in a frame in hyperbolic motion can become frozen to a complete standstill relative to a stationary frame.

(Figures in this article are in colour only in the electronic version.)

## 1. Introduction

This paper discusses some peculiar properties of special relativity involving hyperbolic spacetime motion [1]. In a reference frame in hyperbolic motion relative to a stationary frame, time can be arranged to go arbitrarily slowly [2] in the moving frame compared to the stationary frame. It is a characteristic of hyperbolic motion that this can be achieved even at moderate relativistic velocities. The question then arises whether this slowing-down of time in a frame in hyperbolic motion can be arranged even to freeze time to a complete standstill in the moving frame, and maybe even do so at moderate relativistic velocities. The analysis in this paper seems to answer these questions in the affirmative.

## 2. Hyperbolic motion

Subjecting a test object to a constant acceleration $\alpha$ (as measured in the moving frame $x'y'z'\tau'$ where the test object is at rest) causes the spacetime trajectory of the test object to describe a hyperbola in the stationary frame $xyz\tau$, i.e. if the motion is in the $x$-direction, we get [3]

$$x^2 - \tau^2 = 1/\alpha^2 = x_0^2, \quad (1)$$

where the time $\tau$ and acceleration $\alpha$ are expressed relative to the velocity of light $c$ as

$$\tau = c\,t, \quad \alpha = a/c^2 \quad (2)$$

where $t$ and $a$ are the normal time and acceleration, and where $x_0$ is the distance along the $x$-axis from the origin to the hyperbola (see figure 1).

Consider now the Lorentz transformation relating the stationary frame $xyz\tau$ to the frame $x'y'z'\tau'$, moving with its $x'$-axis with velocity $v$ in the positive direction along the $x$-axis of the $xyz\tau$ frame. This transformation is given as follows [4]

$$x = \frac{x' + \beta\,\tau'}{\sqrt{1-\beta^2}},\; y = y',\; z = z',\; \tau = \frac{\tau' + \beta\,x'}{\sqrt{1-\beta^2}}, \quad (3)$$

or conversely [4] for the coordinates of the moving $x'y'z'\tau'$ frame expressed in the coordinates in the stationary $xyz\tau$ frame,

$$x' = \frac{x - \beta\,\tau}{\sqrt{1-\beta^2}},\; y' = y,\; z' = z,\; \tau' = \frac{\tau - \beta\,x}{\sqrt{1-\beta^2}}, \quad (4)$$

where $\beta$ is defined in terms of the velocity $v$ as

$$\beta = v/c. \quad (5)$$

By eliminating $\beta$ between the expressions for $x$ and $\tau$ in (3) or, equivalently, between $x'$ and $\tau'$ in (4), we get

$$x^2 - \tau^2 = x'^2 - \tau'^2. \quad (6)$$

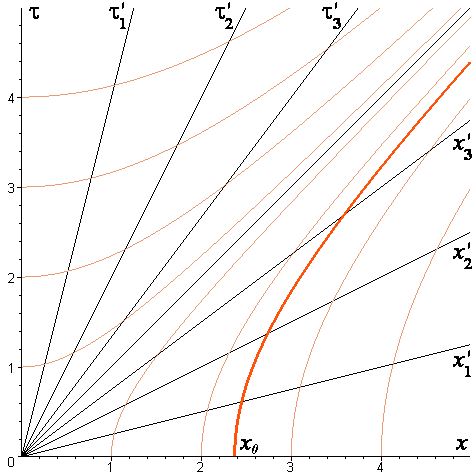

**Figure 1.** Minkowski diagram showing three different positions of the *x'* and *τ'* axes during hyperbolic motion, and with the corresponding scales along the *x'* and *τ'* directions for different velocities shown. Note that the position of a particular coordinate $x' = x_0$, $τ' = 0$ for different velocities describes a hyperbola, just like motion with a corresponding constant acceleration in the co-moving *x'τ'* frame does.

Thus the general world line in the *xyzτ* frame, corresponding to Lorentz transformations of a specific choice of *x'* and *τ'*, is a hyperbola. In particular, the world line in the *xyzτ* frame corresponding to Lorentz transformations of the special choice of $x' = x_0$ and $τ' = 0$ for different *β* is a hyperbola as in (1)

$$x^2 - τ^2 = x_0^2. \qquad (7)$$

### 3. Two coinciding hyperbolas

As will be further discussed in Appendix B, it should be noted that there are actually two coinciding hyperbolic functions involved here: One function is related to the relativistic tilting of the moving frame for different velocities, and is derived above from the Lorentz transformation to be the hyperbola (7). The other function is related to the acceleration of the moving frame relative to the stationary frame, and in the present case with constant acceleration in the moving frame, this other function (1) is also a hyperbola, and can be chosen to be exactly the same hyperbola as (7).

However, it is important to note that hyperbolic motions $x(τ)$, $x'(τ')$ derived from (3) and (4) can be more general than what can be derived from the constant acceleration in the co-moving frame described in section 2. In particular, hyperbolic motions like the solid curve in figure 1 with $x' = x_0$, $τ' = 0$ for different $β(τ)$ exist, this even though such hyperbolic motion cannot be achieved by the constant acceleration in the co-moving frame described in section 2, and which thus gives only a subset of all possible hyperbolic motions. Nevertheless, the resulting hyperbolas corresponding to different such motions $x(τ)$, $x'(τ')$ may still be the same (cf FAQ #2a below).

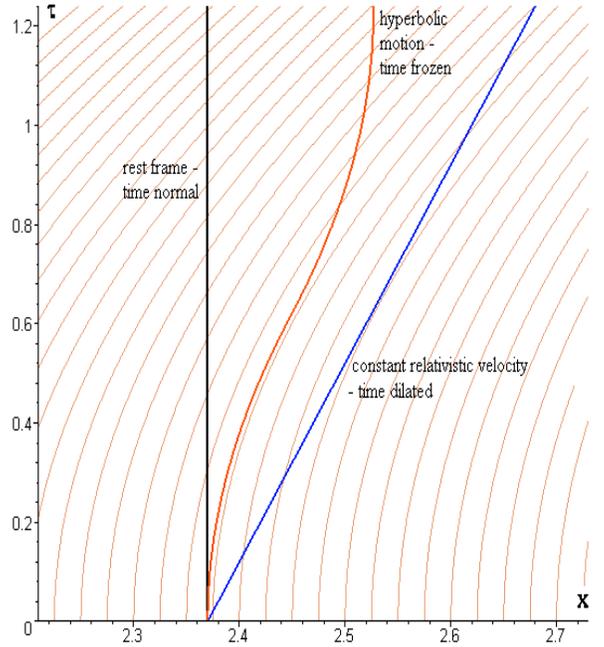

**Figure 2.** Detail of Minkowski diagram as in figure 1 showing (centre curve) an example of hyperbolic acceleration from $β_0 = 0$ to $β_1 ≈ 0.25$, followed by a hyperbolic deceleration back to $β_0 = 0$ again, and taking place at frozen time in the moving frame. For comparison, the figure also shows the world lines for a stationary object (left), and for an object with constant velocity $β_1 ≈ 0.25$ (right) experiencing relativistic time dilation of ca 3 %.

It should also be remarked that even though the *x'τ'*-frame above is accelerating with respect to the *xτ*-frame, the Lorentz transformation can still be used to describe such motion [5]. When discussing hyperbolic motion, the Lorentz transformations leading to the hyperbolas (1) and (7) above can be considered to have taken place between the stationary *xτ*-frame and an infinite sequence of inertial frames *x"τ"*, each one moving with a different constant velocity, and always one momentarily coinciding with the *x'τ'*-frame.

### 4. Frozen time

The above result (7) thus means that the moving frame *x'y'z'τ'* can be chosen in such a way that the time coordinate *τ'* is frozen as $τ' = 0$. The *x'*-coordinate is correspondingly frozen as $x' = x_0$, and moving along the specific hyperbola in the *xyzτ* frame with constant $x' = x_0$, $τ' = 0$ in the *x'y'z'τ'* frame, as shown in figure 1.

Hyperbolic motion is thus interesting in that it can be made to involve a type of motion relative to a stationary reference frame such that time can be frozen in the moving frame compared to the stationary frame. In a science-fiction type context, this would thus open a theoretical possibility for an advanced civilisation to reach distant worlds without the crew ageing at all, despite perhaps centuries passing on their home planet.

In particular it should be noted that this is a real freezing of time in the moving frame, not just the slowing-down of time resulting from the relativistic time dilation due to the Lorentz transformation at velocities very close to the speed of light, which is often discussed in the literature [6]. In fact, it should be observed that the freezing of time discussed above can occur for hyperbolic motion even in a velocity range where the maximum velocity is substantially less than the velocity of light as is illustrated in figure 2.

Figure 2 thus shows a hyperbolic acceleration from rest ($\beta_0 = 0$) to velocity $\beta_1 \approx 0.25$, followed by a hyperbolic deceleration back to rest again. This motion is compared to a rectilinear relativistic motion, in which the object is assumed to be accelerated from rest to the relativistic velocity $\beta_1 \approx 0.25$ in an arbitrarily short time [7], and similarly at the end of the journey quickly decelerated from $\beta_1$ back to rest again. According to Einstein's formula for time dilation, time in that case is retarded by ca 3 %. In contrast to this minute time dilation, time in the moving frame in hyperbolic motion has completely stopped – even though we are here moving at only a moderate fraction of the velocity of light.

We note from (4) that in the special case that the velocity $\beta$ is related to $x$ and $\tau$ in the stationary $xyz\tau$ frame as

$$\beta = \tau / x, \tag{8a}$$

then the time $\tau'$ in the moving $x'y'z'\tau'$ frame would be frozen as $\tau' = 0$ during the motion of this frame relative to the stationary $xyz\tau$ frame. Spacetime motion along a hyperbola as in (7) may thus take place at constant time $\tau'$ in the moving $x'y'z'\tau'$ frame (and with a correspondingly constant $x' = x_0$).

Solving $\tau$ from (8a) and inserting in (4) we get

$$x' = x_0 = x\sqrt{1-\beta^2}, \tag{8b}$$

$$\tau' = 0. \tag{8c}$$

The particular conditions described by (8a) – (8c) can be seen [insert (8a) into (8b) and square both sides] to correspond to the special class of hyperbolas shown in figure 1 with a common asymptote,

$$x^2 - \tau^2 = x_0^2. \tag{9}$$

Other hyperbolas will have other relationships (8a) and will not normally correspond to frozen time (see also the detailed discussion of the examples in Appendix B).

## 5. General vertices

Above we have for simplicity studied hyperbolic motion with the asymptotes $\beta = \pm 1$ intersecting at the origin. In the general case when the asymptotes intersect at an arbitrary vertex $(x_1, \tau_1)$ instead of at the origin, then equations (7), (8a), (8b), and (8c) become, respectively,

$$(x - x_1)^2 - (\tau - \tau_1)^2 = s^2, \tag{10}$$

$$\beta(\tau) = (\tau - \tau_1)/(x - x_1), \tag{11a}$$

$$x' = (x - x_1)\sqrt{1-\beta^2}, \tag{11b}$$

$$\tau' = 0. \tag{11c}$$

*Corresponding to every such vertex $(x_1, \tau_1)$ in the stationary frame, there is thus a specific critical hyperbolic velocity function $\beta = \beta(\tau)$ as defined in (11a) such that the time $\tau'$ in (11c) becomes frozen in the moving frame.*

The crucial point to notice here is thus that corresponding to this particular vertex $(x_1, \tau_1)$, the time $\tau'$ will be frozen as in (11c) *only* for the specific velocity function $\beta(\tau)$ in (11a) – for other hyperbolic velocity functions corresponding to $(x_1, \tau_1)$ we will have a "conventional" hyperbolic motion without frozen time (cf also the discussion in Appendix B).

As discussed earlier in this paper, the condition for the freezing of time is expected to be present even at modest relativistic velocities. However, the special velocity function (11a) is required on $\beta(\tau)$ in order for time to be frozen. This special, unique requirement may thus possibly be a reason why this effect has never happened to be observed in hyperbolic motion at modest relativistic velocities.

## 6. Concluding remarks

The idea that time could be frozen to a complete standstill in an accelerating frame may seem strange. However, it should be remarked that in principle this is nothing essentially new compared to the usual relativistic time-dilation at high velocities (or the slowing-down or even freezing of time in very strong gravitational fields); it is just a different manifestation of essentially the same thing. What is remarkable, however, is that such freezing of time may in this case take place even at more 'normal' conditions and velocities that are substantially lower than the speed of light.

In view of the theoretical (and maybe even potentially practical) implications, it would be highly desirable with further theoretical scrutiny of the results presented in this paper, and if possible experimental verification – or theoretical or experimental refutation of them.

## Appendix A. Alternative derivation of frozen time

Consider the Lorentz transformation (3) relating the stationary frame $x\tau$ to the frame $x'\tau'$, moving with its $x'$-axis with velocity $\beta = v/c$ in the positive direction along the $x$-axis in the $x\tau$ frame,

$$\tau = \frac{\tau' + \beta x'}{\sqrt{1-\beta^2}} \tag{A1}$$

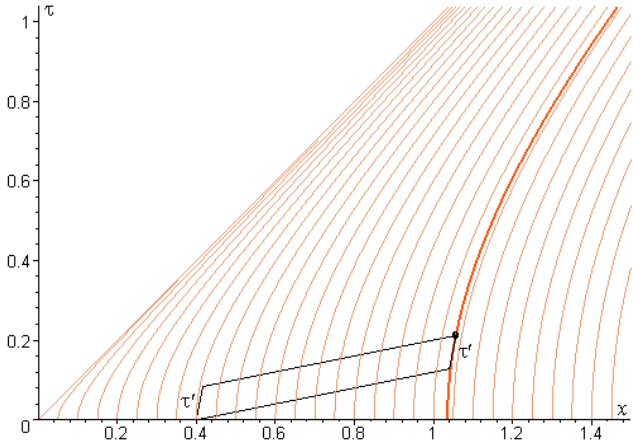

**Figure B1.** Parallelogram showing time $\tau'$ in the moving frame.

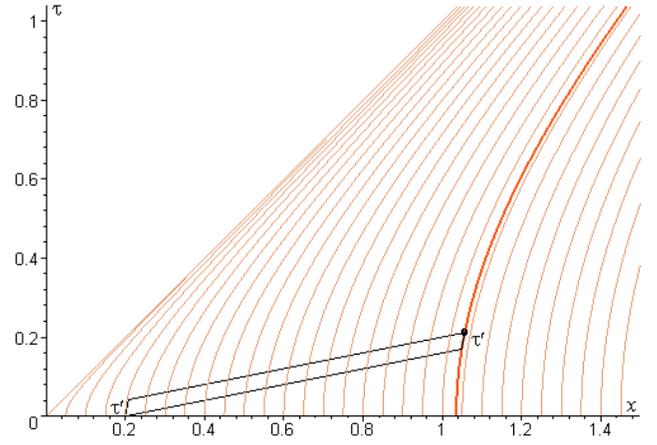

**Figure B2.** As figure B1 but somewhat higher acceleration.

We now calculate the differentials in (A1), for simplicity assuming modest relativistic velocities so that we can use classical addition of velocities for the increment $d\beta$ in the velocity $\beta$, and also so that we can neglect the effect of the increment $d\beta$ in $1-\beta^2$ in the denominator. These two approximations introduce an error $O(\beta^2 d\beta)$ in (A2) below, where we have kept the square root in the denominator intact in order to have an exact expression for all $\beta$ in the case $d\beta = 0$. For a co-moving system we also have $dx' = 0$, and (A1) thus gives

$$d\tau = \frac{d\tau' + x'\,d\beta}{\sqrt{1-\beta^2}} \quad (A2)$$

or

$$d\tau' = d\tau\sqrt{1-\beta^2} - x'\,d\beta \quad (A3)$$

Consider now (cf FAQ #2b) a hyperbolic motion in which the moving frame relative the stationary frame has a constant acceleration $\alpha$ as measured in the moving frame, with $\alpha = a/c^2$ where $a$ is the classical acceleration. The velocity increment $d\beta$ in (A3) due to this acceleration during a time interval $d\tau$ in the stationary frame is then

$$d\beta = \alpha\,d\tau\sqrt{1-\beta^2} \quad (A4)$$

and (A3) thus becomes

$$d\tau' = d\tau\sqrt{1-\beta^2}\,(1 - x'\alpha) \quad (A5)$$

From (A5) we thus see that the time $\tau'$ in the moving frame can be frozen as $d\tau' = 0$ for modest $\beta$, and that this happens if the acceleration $\alpha$ is

$$\alpha = 1/x' \quad (A6)$$

The result for hyperbolic motion at modest relativistic velocities as discussed here is thus in agreement with the general hyperbolic motion in (1) in section 1 (with $x' = x_0$). The alternative, independent derivation of frozen time in (A5) above for moderate relativistic velocities thus supports the derivation of frozen time in (8c) in section 4 earlier in this paper.

**Appendix B. Illustration of frozen time**

In this paper the idea is put forward that time in one frame can be frozen to a complete standstill in comparison to another frame, and could be so even at modest relativistic velocities. If correct, this would seem to be a consequence of relativity that must have been noticed long before. There is thus a considerable burden of proof here, notwithstanding the comment made above in the last paragraph in section 5 that the effect requires special conditions to be met, and for this reason may have escaped accidental discovery.

As a further illustration of the possible method to freeze time as outlined in this paper, a simple example using Minkowski diagrams will in conclusion now be discussed. Three cases of hyperbolic motion as illustrated in figures B1, B2, and B3 will be considered. These cases show how frozen time can be regarded as a special case of a "normal" accelerated motion of hyperbolic type.

The solid curve in figure B1 shows the spacetime trajectory of an object in hyperbolic motion (here with a background of hyperbolas representing different $x$-intercepts within the same light-cone). From a point corresponding to velocity $\beta = 0.2$ (and $x \approx 1.06$, $\tau \approx 0.21$), a parallelogram is drawn, where the slightly tilted short sides represent the time-coordinate $\tau'$ in the moving frame of the object. We see that the relativistic tilting is here low enough to be overtaken by the motion of the object (remember that we actually have two

Figure B3, finally, shows the limiting case when the acceleration of the object is high enough to cause the continuous relativistic tilting of the *x'*-axis during the acceleration to exactly keep pace with the motion of the object. This is also the final case, since the parallelogram has now reached the light-cone, beyond which it can go no further in this way. We see that the parallelogram has now degenerated to a double line – the time $\tau'$ in the moving frame is now frozen.

It should be remarked that the effect discussed here is valid for all the other hyperbolas plotted in the figure, and also for all those hyperbolas not shown far to the right, which may involve arbitrarily small accelerations. Hence it is not an effect requiring critically large accelerations.

### Acknowledgments

The author wishes to thank Dr Hans-Olov Zetterström for many fruitful discussions and a number of relevant and valuable questions and suggestions. My heartfelt thanks also go to Professor Øyvind Grøn for constructive comments and encouragement. I am also most grateful to all the many readers who have critically studied the successive updates of my manuscript and asked all the relevant questions that are now summarised in the collection of FAQ filed in arXiv:1108.1112 [physics.gen-ph] on http://arXiv.org/.

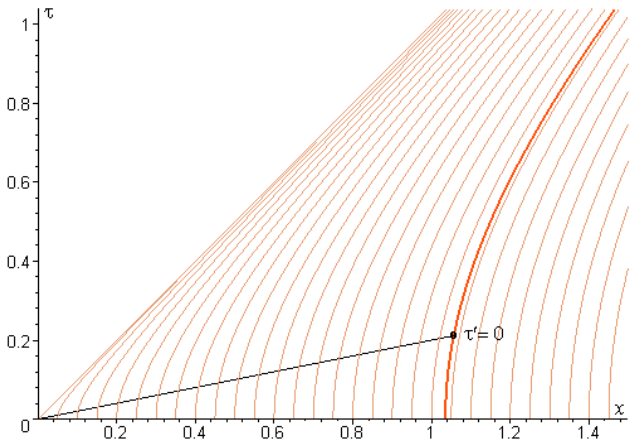

**Figure B3.** Limiting case with time frozen to standstill.

coinciding hyperbolas here, as discussed in the first paragraph of section 3). The continuous relativistic tilting of the *x'*-axis during the acceleration is thus here not sufficient to keep pace with the acceleration of the object. Even though the Lorentz transformation makes time go more slowly in the moving *x'τ'* frame, the time $\tau'$ is thus still passing in this frame.

Figure B2 envisages an object moving along the same hyperbola, but now with a somewhat higher acceleration. This thus means that it will here move faster than in the previous example along the hyperbola representing increasing values of the velocity $\beta$ as seen in the *xτ* frame. We see that the continuously increased relativistic tilting of the *x'*-axis during the acceleration of the object is now beginning to catch up with the motion of the object. However, some time $\tau'$ still passes in the *x'τ'* frame during the motion of the object along the hyperbola.

### References

[1] Pauli W 1981 *Theory of Relativity* (Dover), p 75
[2] Misner C, Thorne K and Wheeler J 1973, *Gravitation* (Freeman), Exercise 6.3 (b), p 167
[3] Møller C 1982, *The Theory of Relativity*, 2nd ed. (Dehli) p 74
[4] Pauli W 1981 *Theory of Relativity* (Dover) p 10
[5] Misner C, Thorne K and Wheeler J 1973, *Gravitation* (Freeman), Ch 6
[6] Tolman R 1987 *Relativity, Thermodynamics and Cosmology* (Dover) p 22
[7] Aharoni J 1985 *The Special Theory of Relativity* (Dover) p 21

# $\underline{\textbf{F}_{\text{requently}}\ \textbf{A}_{\text{sked}}\ \textbf{Q}_{\text{uestions}}}$

(including some questions that should have been asked, but haven't)

**on "Frozen time in hyperbolic spacetime motion"**
**(arXiv:1108.1112) by Arne Bergstrom**

---

## FAQ #1a (page 1)

"**The expression for a world line x($t$) in the stationary frame of a particle subjected to a constant acceleration $a$ in the moving frame is given as follows (G Barton, "Introduction to the Relativity Principle", p 95)**

$$x(t) = \frac{c^2\left(\sqrt{1 + \frac{a^2 t^2}{c^2}} - 1\right)}{a}$$

**How does this tally with your formula (1) on page 1?**"

ANSWER: Consider a motion as given in your question with a constant non-zero acceleration $a$, so that we can freely change variables as follows

$$a = \alpha\, c^2,\ x(t) = x - \frac{1}{\alpha},\ t = \frac{\tau}{c} \tag{1}$$

and we thus get

$$x - \frac{1}{\alpha} = \frac{\sqrt{1 + \alpha^2 \tau^2} - 1}{\alpha} \tag{2}$$

*i e*

$$x = \frac{\sqrt{1 + \alpha^2 \tau^2}}{\alpha} \tag{3}$$

or

$$x^2 = \frac{1}{\alpha^2} + \tau^2 \tag{4}$$

and finally
$$x^2 - \tau^2 = \frac{1}{\alpha^2} \qquad (5)$$

which thus agrees with (1) in my paper (cf also the derivation in ref [3] in my paper).

Please note that $x_0$ in my manuscript is defined as "the distance along the *x*-axis from the origin to the hyperbola" and can thus in my case be set = 0 only in the case the hyperbola is degenerated to its asymptotes.

---

## FAQ #1b (page 1)

"**In a Minkowski diagram, consider a sequence of spacetime frames with a common origin as in Fig. 1 in your paper and which, one after the other, correspond to different velocities that increase in a specific way. Suppose that in the limit of an infinite such sequence, the successive Lorentz-transformations of a specific point $(x_0, 0)$ in them will form a spacetime trajectory in the form of a hyperbola as in your Fig. 1. But where is the motion?**"

ANSWER: Good question! The hyperbola you are discussing corresponds to (7) in the paper. As I say in the second paragraph on page 2 in the paper, the answer to your question is that in the paper I am at the same time actually considering also another, coinciding identical hyperbola, namely (1), which is the spacetime trajectory of a point in constant acceleration (as measured in its own rest frame). Such a point describes a spacetime trajectory that happens also to be a hyperbola, and which with proper choice of acceleration is in fact an identical hyperbola as the one in (7) that you discuss in your question. But the rate of tilting of the coordinate frames as function of time, which gives the first hyperbola, is not related to the motion of the accelerating point as function of time, which gives the second hyperbola. These two accelerations are independent of each other, as described in Appendix B in my paper (see also FAQ #4b below). It is this difference that can lead to a freezing of time as discussed in Appendix B (*cf* FAQ #4c below).

---

## FAQ #2a (page 2)

"**The relationship between rest-frame time $\tau$ and proper time $\tau'$ for a particle subjected to a constant proper acceleration $\alpha$ can be derived to be as follows (expressed in the symbols and units you use)**
$$\tau = \sinh(\alpha \tau') / \alpha \qquad (i)$$

**(cf G Barton, "Introduction to the Relativity Principle" 1999, p 97).**

**If so, how can the proper time $\tau'$ be frozen as you claim in your paper?**"

ANSWER: It is important to realise that hyperbolic motion in accordance with (1) and (7) in the paper is more general than what can be achieved by the particular constant acceleration $\alpha$ in the co-moving frame that you discuss in your question. Your expression (*i*) above only gives a subset of all possible hyperbolic trajectories $x(\tau)$, $x'(\tau')$, and specifically results in the following relationships between the coordinates $x\tau$ in the stationary frame and the coordinates $x'\tau'$ in the co-moving frame (see Misner C, Thorne K and Wheeler J 1973, *Gravitation*, p 166)

$$x = \cosh(\alpha\tau')/\alpha; \quad \tau = \sinh(\alpha\tau')/\alpha, \tag{1}$$

giving the following hyperbola

$$x^2 - \tau^2 = 1/\alpha^2, \tag{2}$$

It is important to note that spacetime motion according to (1) above is not equivalent to hyperbolic motion: The motion defined by (1) above implies the hyperbolic motion in (2), but the hyperbolic motion in (2) does not necessarily imply a motion according to (1) above. Thus (1) above is only one example of possible relationships $x(\tau')$, $\tau(\tau')$ between the coordinates $x\tau$ in the stationary frame and the coordinates $x'\tau'$ in the co-moving frame that result in the hyperbolic motion in (2). But the relationship in (1) does not, *e g*, describe the red legitimate hyperbolic trajectory in figure 1 in the paper. Another, more general relationship – and which does so – would be as follows

$$x = \cosh(f(\tau'))/\alpha; \quad \tau = \sinh(f(\tau'))/\alpha, \tag{3}$$

which for $f(\tau') \neq \alpha'\tau'$ would obviously not correspond to the motion discussed in (1) above, but would nevertheless give the same hyperbola as in (2)

$$x^2 - \tau^2 = 1/\alpha^2.$$

In summary, hyperbolic motion is thus more general than just the result of the specific constant proper acceleration given in (*i*) and (1) above. Time may thus be frozen in hyperbolic motion regardless of restrictions seemingly imposed by reference to results derived for the particular constant proper acceleration given in your question.

*Additional comment 1:* The above properties of hyperbolic motion can be further discussed as follows. With a more general hyperbolic motion as in (3) above, we have the following relationships between the coordinates $x\tau$ in the stationary frame and the coordinates $x'\tau'$ in the co-moving frame

$$x = \cosh(f(\tau'))/\alpha, \tag{4}$$

$$\tau = \sinh(f(\tau'))/\alpha, \tag{5}$$

which correspond to a hyperbola

$$x^2 - \tau^2 = \cosh(f(\tau'))^2/\alpha^2 - \sinh(f(\tau'))^2/\alpha^2 = 1/\alpha^2. \tag{6}$$

Differentiating (4) and (5) we get the velocity in the stationary frame

$$\beta = \tanh(f(\tau')) \tag{7}$$

and thus also

$$\gamma = \frac{1}{\sqrt{1-\beta^2}} = \cosh(f(\tau'))\tag{8}$$

Differentiating (7) and (5) with respect to $\tau'$ in order to calculate $\partial\beta/\partial\tau'$ and $\partial\tau/\partial\tau'$, and using (8), we then get for the acceleration in the stationary frame

$$\frac{\partial}{\partial\tau}\beta = \frac{\alpha}{\gamma^3}\tag{9}$$

which transforms (G Barton, "Introduction to the Relativity Principle", p 94) to an acceleration $\alpha$ in the co-moving frame. Note that the quantity $\alpha$ in (3) above [and maybe even in the denominator in (1) above] has the character of a scale factor rather than an acceleration. For this reason, it could perhaps be natural that exactly the same factor $\alpha$ appears when we calculate the proper acceleration from several different types of hyperbolic motion, as it does above from both (1) and (3). Those different variants of hyperbolic motion may then contain variants that do not allow frozen time [like (1) above] as well as more general variants [like (3) above], which do allow frozen time. These variants correspond to quite different behavior of the acceleration in the stationary frame as seen in Fig. 5 below – but nevertheless give the same proper acceleration as also seen in Fig. 5.

Figs. 1 – 5 below illustrate the trajectories, velocities and accelerations derived from (1) and (3), respectively.

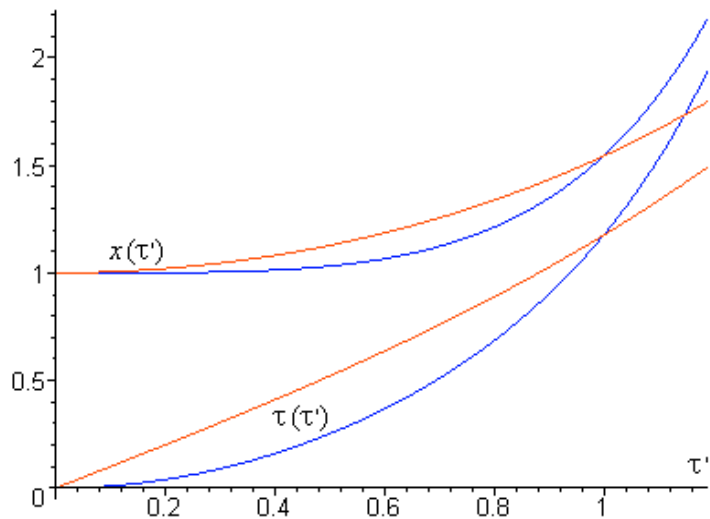

Fig 1. Examples of space coordinates $x(\tau')$ [upper curves] and time coordinates $\tau(\tau')$ [lower curves] in (3) above as functions of $\tau'$ for, respectively, $f(\tau') = \tau'$ (red) and $f(\tau') = \tau'^2$ (blue). The red case corresponds to the 'traditional' hyperbolic motion in (1) with approximately constant acceleration even in the stationary frame at low velocities.

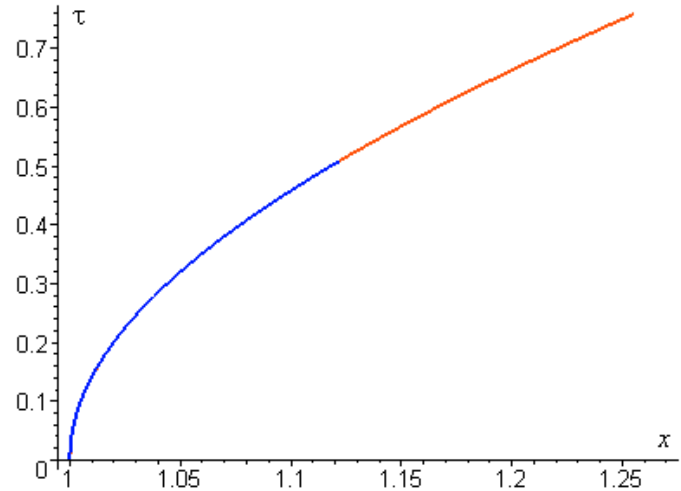

Fig 2. Time coordinate $\tau(\tau')$ in (5) as function of space coordinate $x(\tau')$ in (4) for $f(\tau') = \tau'$ (red) and $f(\tau') = \tau'^2$ (blue) [cf Fig. 1]. Note that both curves form overlapping hyperbolas, but that the blue hyperbola moves more slowly than the red one when $\tau'$ increases (both curves show the interval $0 < \tau' < 0.7$).

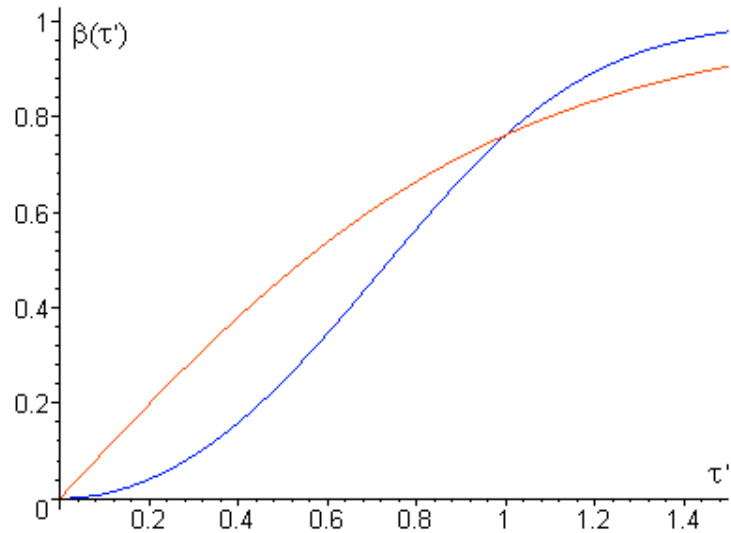

Fig 3. Velocities $\beta(\tau')$ in (7) in the stationary frame as function of $\tau'$ corresponding to the red and blue cases in Fig. 1. Note how the red curve corresponding to $f(\tau') = \tau'$ starts off as a straight line as expected for a constant acceleration and then bends down somewhat due to relativistic effects. In contrast, the blue curve corresponding to $f(\tau') = \tau'^2$ has a more unconventional dependence of time $\tau'$ since $f(\tau')$ is quadratic in $\tau'$ (but it is still a legitimate motion and gives the same proper acceleration as seen in Fig. 5).

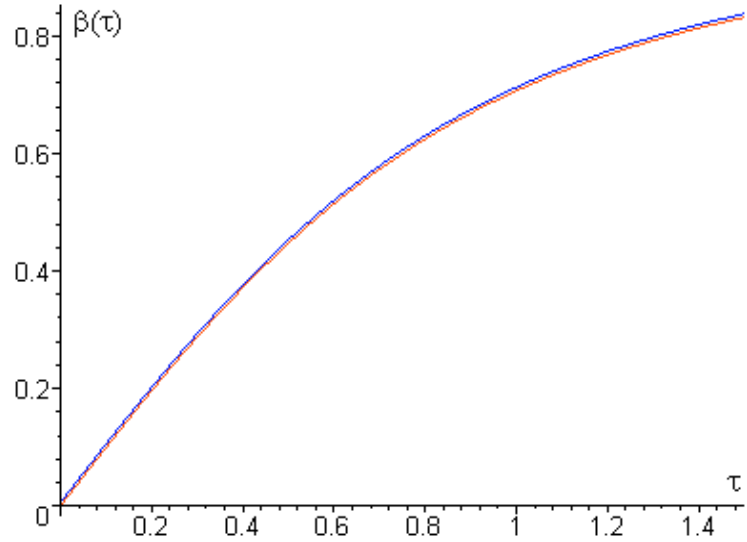

Fig 4. Velocities $\beta(\tau)$ in the stationary frame as function of $\tau$ corresponding to the red and blue cases in Fig. 1. Note how both the red curve corresponding to $f(\tau') = \tau'$ and the blue curve corresponding to $f(\tau') = \tau'^2$ in the co-moving frame coincide since in both cases we have from (7) and (5) that $\beta(\tau) = \tanh(f(\tau')) = \sinh(f(\tau'))/\sqrt{1 + \sinh(f(\tau'))^2} = \tau/\sqrt{1 + \tau^2}$.

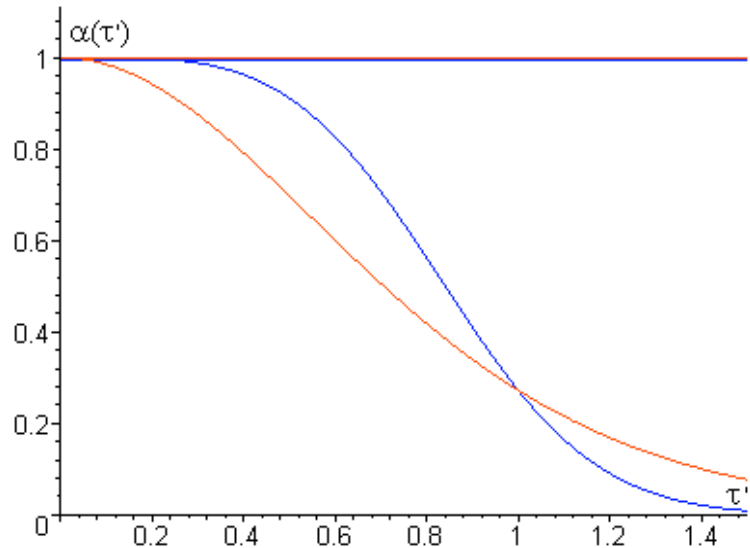

Fig 5. Accelerations $\alpha(\tau')$ as seen in the stationary frame according to eq (9) (lower two curves) and as seen in the co-moving frame (upper two coinciding straight lines). Note that in the co-moving frame, both accelerations are equal and constant as discussed after (9) on page 9 above.

With the function $f(\tau')$ as defined in (3) above, the momentary acceleration in the co-moving frame is thus always α. We can thus have a hyperbolic motion as in (2) above even in other cases than derived as in (1) above. Thus, for instance, the red hyperbola in Fig. 1 in the paper with $f(\tau') = 0$ and $x' = x_0$, $\tau' = 0$ represents a legitimate hyperbolic motion. *As stated above, hyperbolic motion is thus more general than just the result of the particular constant proper acceleration leading to (i) and (1) above.*

*Additional comment 2:* For a constant relativistic acceleration $\alpha = a/c^2$ in the stationary frame, the time $\tau'$ elapsed in the co-moving frame is given as follows (cf G Barton, "Introduction to the Relativity Principle", p 95)

$$\tau' = \frac{\ln(\alpha\tau + \sqrt{1 + \alpha^2 \tau^2})}{\alpha} \tag{11}$$

or equivalently as in (1),

$$\tau = \sinh(\alpha\tau')/\alpha. \tag{1}$$

Again, this can be used to prove that the proper time $\tau'$ cannot be frozen for systems in hyperbolic motion due to a constant proper acceleration as in (1) above. *But that is not the same as saying that the proper time $\tau'$ cannot be frozen for hyperbolic motion, which is the subject matter here.* As shown in the comment above, hyperbolic motion is more general than the motion just discussed, which involves the particular constant proper acceleration in (1) above. We can thus have frozen time in hyperbolic motions involving systems that are subjected to other types of accelerations as exemplified by the more general expression in (3) above.

### Summary:

**As illustrated in the above figures, we see that the general hyperbolic trajectory in (3) on page 8 gives the same spacetime trajectory $x(\tau)$ [Fig. 2], velocity $\beta(\tau)$ [Fig. 4] and constant proper acceleration $\alpha(\tau)$ [upper straight lines in Fig. 5] as the 'traditional' hyperbolic trajectory in (1), but compared to which the general hyperbolic trajectory in (3) permits frozen time in the proper frame.**

---

## FAQ #2b (page 2)

"**The relationship between a time interval $d\tau'$ in a frame moving with velocity $\beta = v/c$ and the corresponding time interval $d\tau$ in a stationary frame is given as follows ("the clock hypothesis")**

$$d\tau' = d\tau\sqrt{1 - \beta^2} \tag{i}$$

**If so, how can the time $\tau'$ be frozen for any system, accelerating or not, for velocities substantially lower than the velocity of light as you claim in your paper?**"

ANSWER: It should be remarked that the so-called "clock hypothesis" (i) above was never any part of Einstein's theory, but is a later addition (Wikipedia, "Clock hypothesis"), and which may actually not be strictly correct for accelerated frames. We need to remember that the Lorentz transformation involves both the space and time coordinates, and this fact is important when we are considering accelerated motion, as will be seen from equation (6) below. In principle, what can be seen to happen is that when we calculate the relationship between $\tau$ and $\tau'$ by the Lorentz transformation, we also have a $x'$-term. If we then form the differential $d\tau$, then – in the case when $\beta$ is constant – the $x'$-terms between one time-step and the next cancel, and we get an expression as in your (i). But only if $\beta$ is constant. If $\beta$ is not constant, then the contributions from the $x'$-terms between one time-step and the next do not cancel any longer – since there is now a factor $\beta$ in the $x'$-terms, and this factor is then different between the time steps. (It should also be remarked that this behaviour is independent of the conceived replacement of the accelerated frame by a succession of inertial frames discussed in Sect. 3 in the paper, since there will still be the difference just discussed in the velocity between successive frames.) Below is a detailed derivation as outlined above.

Consider, as in (3) in the paper, the Lorentz transformation relating the stationary frame $x\tau$ to the frame $x'\tau'$, moving with its $x'$-axis with velocity $\beta = v/c$ in the positive direction along the $x$-axis in the $x\tau$ frame. This transformation is given as follows

$$\tau = \frac{\tau' + \beta x'}{\sqrt{1-\beta^2}} \tag{1}$$

Now consider this expression when we make the increments $\tau \rightarrow \tau + d\tau$, $\tau' \rightarrow \tau' + d\tau'$, $x' \rightarrow x' + dx'$, $\beta \rightarrow \beta + d\beta$. For maximum transparency we consider here for simplicity the case of modest relativistic velocities so that we can (*a*) use classical addition of velocities in $\beta + d\beta$, and also (*b*) neglect the effect of the increment of $\beta$ in the denominator. The approximations (*a*) and (*b*) introduce an error $O(\beta^2 d\beta)$ in (2) below, where we have kept the square root in the denominator intact in order to have an exact expression for all $\beta$ in the case $d\beta = 0$. With the above increments and assumptions, (1) thus becomes

$$\tau + d\tau = \frac{\tau' + d\tau' + (\beta + d\beta)(x' + dx')}{\sqrt{1-\beta^2}} \tag{2}$$

For a co-moving system we have $dx' = 0$. Subtracting (1) from (2), we then get

$$d\tau = \frac{d\tau' + x' d\beta}{\sqrt{1-\beta^2}} \tag{3}$$

or (note the $x'$-term)

$$d\tau' = d\tau \sqrt{1-\beta^2} - x' d\beta \tag{4}$$

Consider now hyperbolic motion in which the moving frame relative the stationary frame has a constant acceleration $\alpha$ as measured in the moving frame, with $\alpha = a/c^2$ where $a$ is the classical acceleration. The velocity increment $d\beta$ due to this acceleration during a time interval $d\tau$ in the stationary frame is

$$d\beta = \alpha \, d\tau \sqrt{1-\beta^2} \tag{5}$$

and (4) thus becomes [to order $O(\beta^2 d\beta)$]

$$d\tau' = d\tau \sqrt{1-\beta^2} \, (1 - x'\alpha) \tag{6}$$

From (6) we thus see that the time $\tau'$ in the moving system for this hyperbolic motion can indeed be frozen as $d\tau' = 0$ for modest $\beta$, and that this happens if the acceleration $\alpha$ is

$$\alpha = 1/x'. \qquad (7)$$

The hyperbolic motion we are studying here thus agrees (set $x' = x_0$) with the hyperbolic motion derived earlier in FAQ #1a and with the hyperbolic motion derived in my paper. The present result is thus an independent verification of the derivation of frozen time in Sect. 4 in my paper (although it is here in this FAQ for simplicity derived only for modest relativistic velocities).

---

# FAQ #3a (page 3)

"**Can you elaborate on the difference between the traditional relativistic time dilation and the slowing-down and possible freezing of time you discuss in your paper?**"

ANSWER: The traditional relativistic time dilation can be calculated from the Lorentz transformation (4) in the paper. The time $\tau'$ in a frame moving with (constant) velocity $\beta = v/c$ is then related to time $\tau$ in the stationary frame as

$$\tau' = \tau / \sqrt{1 - \beta^2}$$

Inserting $\beta = 0.25$ as in Fig. 2 in the paper, we get $\tau' = \tau / \sqrt{(1 - 0.25^2)} \approx 1.033\ \tau$, i e time in the moving frame goes about 3 % slower than in the stationary frame, as I state on page 2 in my paper.

In contrast to this traditional relativistic time dilation occurring at (constant) velocities close to the speed of light, there is also a time dilation experienced by an object in hyperbolic motion as discussed in my paper, and illustrated in Figs. B1 – B3 in the paper (and also in Figs. I – III in FAQ #4c below). This time dilation is of quite different nature and an effect of the acceleration in the hyperbolic motion. It differs from the traditional relativistic time dilation in the following two important respects:

(1) it can cause an appreciable slowing-down of time even at relativistic velocities substantially lower than the velocity of light, and

(2) it can be arranged not just to slow down time, but also actually to freeze time to a complete standstill, and do so at relativistic velocities substantially lower than the velocity of light.

---

# FAQ #4a (pages 4 - 5)

"**How are the parallelograms in Figs. B1 and B2 on pages 3 and 4 constructed?**"

ANSWER: Select an arbitrary point on the hyperbola as the topmost point $O$ of the parallelogram (marked with a black dot in the figures in the paper). From this point draw the tangent to the hyperbola downwards. From an arbitrary point, P say, somewhere on this tangent (which thus becomes the first side OP in the parallelogram), draw a 45-degree help-line (light-ray) towards the upper left, and measure the angle Ω between the tangent and this help-line at the point P. Draw the second side in the parallelogram from P so that it forms the same angle Ω on the other side of the help-line (i e so that the second side PQ in the parallelogram forms the same angle with the horizontal x-axis as the first side OP forms with the vertical τ-axis). The second side of the parallelogram intersects the x-axis in the point Q.

Now finish the parallelogram by drawing the third side QR to R parallel to the first side OP (and of equal length as the first side), and then draw the fourth side RQ parallel to the second side PQ (and of equal length as the second side). This closes the parallelogram.

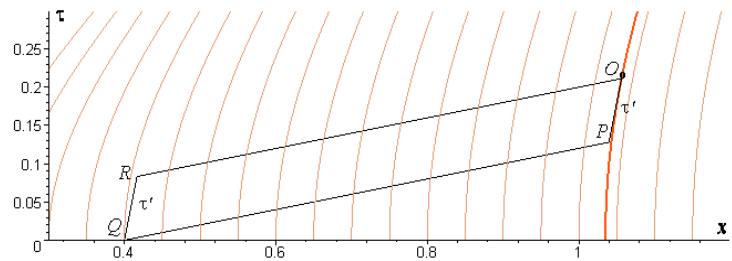

Note that in this construction the choice of the point P is arbitrary. This freedom in the choice of the point P is thus what gives the difference between Figs. B1 and B2 with different times τ' – and also creates the special case in Fig. B3, in which time is frozen.

Alternatively (and as is actually done in the figures in the paper), one can construct the parallelogram as follows: After having selected the arbitrary topmost point O on the hyperbola, choose an arbitrary x-intercept Q. Calculate the derivative $d_1$ of the hyperbola in O and write the equation for a line $L_1$ with this slope $d_1$ through O. Then invert this derivative, $1/d_1 = d_2$, and write the equation for a line $L_2$ with that slope $d_2$ through Q. Solve the simultaneous equations $L_1$ and $L_2$ to get the coordinates for the point P. Finally, write the equation for a line $L_3$ through Q with slope $d_1$, and for a line $L_4$ through O with slope $d_2$, and solve the simultaneous equations $L_3$ and $L_4$ to get the coordinates for the point R.

Note that again there is a freedom in the construction, in this case in the choice of the x-intercept Q, thus giving the difference between Figs. B1 and B2 with different times τ' – and also making possible the special case in Fig. B3, in which time is frozen.

---

## FAQ #4b (page 4)

**"It seems from your answer above as if the relativistic tilting of the moving frame is like in competition with the motion of the object along the hyperbola."**

ANSWER: Exactly! As I discuss in FAQ #1b above, there are actually two hyperbolic functions involved: The first function is related to the relativistic tilting of the moving frame for different velocities, and is derived from the Lorentz transformation to be a hyperbola. The second function (which in the general case need not be a hyperbolic function at all) is related to the acceleration of the moving frame relative to the stationary frame. In this particular case with constant acceleration in the moving frame, the second function is also a hyperbola and can be chosen to be exactly the same hyperbolic function as the first function. However, even in that case their respective scalings $\tau(\beta)$ may differ. We normally have a situation when the relativistic acceleration of the object dominates over the speed in which the relativistic tilting of the spacetime frame takes place. A passage of time is then observed in the moving frame, albeit clocks move more slowly than in the stationary frame. But in the limit when the *x*-intercept $Q$ in the figure in FAQ #4a reaches the origin, then time grinds to a complete halt in the moving frame as observed from the stationary frame (cf Fig B3 in the paper and Fig. III in FAQ #4c below).

The important thing to note here is that this is a completely different kind of time dilation from the one occurring very close to the speed of light according to the Lorentz transformation. In this case with time dilation due to hyperbolic motion, time may not only slow down much more for moderate velocities than in the traditional time dilation, but it can actually even be brought to a complete standstill in the moving frame – and may do so even at modest relativistic velocities.

---

## FAQ #4c (pages 4-5)

"**For comparison, it would be interesting to see the parallelograms in Figs. B1 - B3 in your paper plotted for some other value of $\beta$ as well. In particular, it would be interesting to see that the limiting case (when the parallelogram degenerates to a double line) actually then occurs for all values of $\beta$.**"

ANSWER: In Figs. I – III below, the corresponding parallelograms as for $\beta = 0.2$ in Figs. B1 – B3 in the paper are plotted also for $\beta = 0.6$ (upper parallelograms). Again we see how the parallelograms in the final picture (Fig. III) have degenerated to double lines corresponding to frozen time.

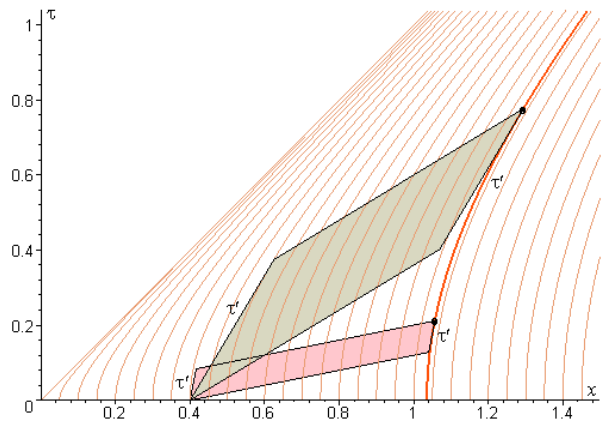

Fig. I. *x*-intercept = 0.4

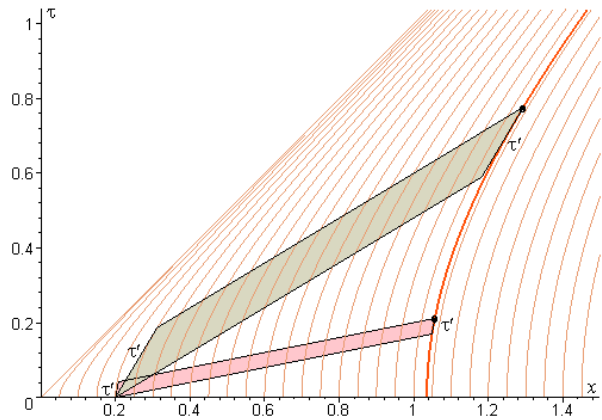

Fig. II. *x*-intercept = 0.2

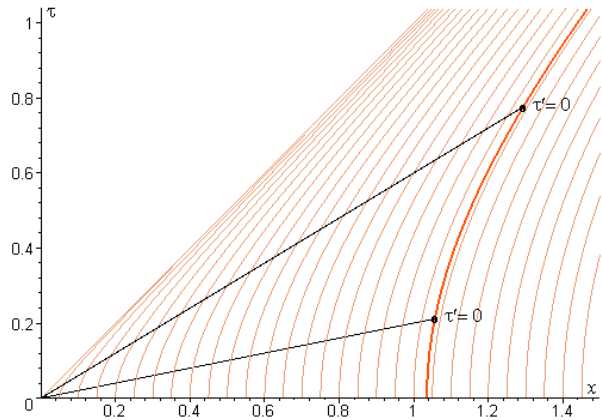

Fig. III. *x*-intercept = 0

The figures show the positions (black dots) for the same two velocities $\beta = 0.2$ and $\beta = 0.6$ in each of the three cases. However, the accelerations of the moving frame as expressed in the time-coordinate $\tau$ are different for each of the three cases. The acceleration of the moving frame is smallest in Fig. I, which means that in this figure it takes longer time for the points marked as black dots to move along the hyperbola; the side $\tau'$ is longer in this case than in the following figures.

In Fig. II the points marked as black dots move faster along the different velocities $\beta$ on the hyperbola; the side $\tau'$ is shorter in this case than in the preceding case in Fig. I.

In Fig. III, finally, the side $\tau'$ has shrunk to zero – corresponding to frozen time – and this obviously happens for all velocities $\beta$ along the hyperbola. This would thus permit freezing of time in a moving frame in hyperbolic motion relative to a stationary frame even at modest relativistic velocities.

---

23 August 2011

Arne Bergstrom